\magnification = 1200
\baselineskip 20 pt
\nopagenumbers
\voffset 1.25 in
\centerline {\bf A family of exact solutions for}

\centerline {\bf unpolarized Gowdy models}
\vskip 20 pt

\centerline {Octavio Obreg\'on}
\vskip 10 pt

\centerline {\sl Instituto de F{\'\i}sica, Universidad de Guanajuato}

\centerline {\sl A. Postal E-143}

\centerline {\sl Le\'on, Guanajuato, MEXICO}
\vskip 10 pt

\centerline {and}
\vskip 10 pt

\centerline {Michael P. Ryan, Jr.}
\vskip 10 pt

\centerline {\sl Instituto de Ciencias Nucleares, Universidad Nacional
Aut\'onoma de M\'exico}

\centerline {\sl A. Postal 70-543}

\centerline {\sl M\'exico 04510 D.F., MEXICO}
\vskip 10 pt

\vfill\eject

\voffset 0 in
\baselineskip = 10 pt
\noindent ABSTRACT
\vskip 10 pt

Unpolarized Gowdy models are inhomogeneous cosmological models that depend
on time and one spatial variable with complicated nonlinear equations
of motion.  There are two topologies associated with these models, $T^3$
(three-torus) and $S^1 \times S^2$.  The $T^3$ models have been used for
numerical studies because it seems to be difficult to find analytic
solutions to their nonlinear Einstein equations.  The $S^1 \times S^2$
models have even more complicated equations, but at least one family of
analytic solutions can be given as a reinterpretation of known solutions.
Various properties of this family of solutions are studied.
\vfill\eject

\pageno = 1
\footline{\hss\tenrm\folio\hss}

\noindent {\bf I. Introduction}
\vskip 10 pt

In 1971 Gowdy proposed a set of inhomogeneous cosmological metrics with the
idea of constructing vacuum models in which inhomogeneity appeared in a
simple way [1].  He wanted his models to have compact topology and two
commuting Killing vectors constructed so that the model depended only on
time and {\it one\/} of the spatial variables.  Models of this sort have
the metric structure of axisymmetric or cylindrically symmetric static
or stationary metrics with the radial coordinate replaced by $t$ and the
metric independent of the new ``radial'' coordinate instead of being
independent of time.  Such metrics share many of the properties of
axisymmetric metrics and a number of well-known results for these metrics
can be applied to the Gowdy models almost without change.

There are two topologies for the $t$ = constant three surfaces of these
metrics that were proposed by Gowdy.  One was a $T^3$ (three-torus) topology
and another was an $S^1 \times S^2$ topology.  For some reason the $T^3$
topology metrics have been used extensively for a number of studies of
different types, while the models with $S^1 \times S^2$ topology have
been almost forgotten.

In the notation of Berger and Garfinkle [2] the original $T^3$ Gowdy
model takes the form
$$ds^2 = e^{-\lambda /2} e^{\tau /2} (-e^{-2\tau} d\tau^2 + d\theta^2) +
e^{-\tau}[e^P d\sigma^2 + e^{-P} d\delta^2], \eqno (1.1)$$
where $\lambda$ and $P$ are functions of $\theta$ and $\tau$.  The model
is compactified by requiring $0 \leq \theta , \sigma , \delta \leq 2\pi$,
and the metric functions must be periodic in $\theta$.  It is often
stated that this form of the metric can be achieved by coordinate
transformations alone, but (see for example, [3]) the form of the
coefficient of $d\delta^2$ comes from the fact that the determinant of the
two-metric $g_2$ formed by the $d\sigma^2$ and $d\delta^2$ terms, which in
this case is $e^{-2\tau}$, is a solution of a certain combination of the
Einstein equations.  If we assume that $g_2$ is $g_2 (\tau, \theta)$, and
define $T \equiv e^{-\tau}$, then
$$R^{\delta}_{\delta} + R^{\sigma}_{\sigma} = 0 \Rightarrow
{{(g_2)_{,TT}}\over
{g_2}} - {{1}\over {2}}\left ( {{[g_2]_{,T}}\over {g_2}} \right )^2 -
\left [ {{(g_2)_{,\theta \theta}}\over {g_2}} - {{1}\over {2}} \left (
{{[g_2]_{,\theta}}\over {g_2}} \right )^2 \right ] = 0. \eqno (1.2)$$
This expression is valid for any two-metric analogous to the $d\sigma^2$,
$d\delta^2$ part of (1.1) of the form $A(\theta ,\tau) d\sigma^2 +
2B(\theta , \tau) d\sigma d\delta + C(\theta , \tau) d\delta^2$, not just
the special case given in (1.1), a fact that we will use below.  In the
case of (1.1) $g_2 = T^2 = e^{-2\tau}$ is a solution of (1.2), which
allows us to write the $d\delta^2$ term of (1.1) in the form given above.

The Einstein equations for (1.1) become a linear equation for $P$ and a
set of first-order equations for $\lambda$ which can be integrated to
give $\lambda$ given any solution to the equation for $P$.  These
Einstein equations are easily solved in terms of ordinary Bessel functions,
and several particular solutions were given by Gowdy [1].

Gowdy [4] also proposed ``unpolarized'' models (the name comes from the
fact that for small values of $P$ the metric [1.1] becomes that of polarized
linearized gravitational waves, and the unpolarized models become that of
unpolarized waves), which, in the notation of [2] become, for the $T^3$
models,
$$ds^2 = e^{-\lambda /2} e^{\tau /2} (-e^{-2\tau} d\tau^2 + d\theta^2)$$
$$+ e^{-\tau} (e^P d\sigma^2 + 2e^P Q d\sigma d\delta + [e^{-P} + e^P
Q^2]d\delta^2), \eqno (1.3)$$
where the form of the $d\delta^2$ term comes from taking $g_2 = e^{-2\tau}$
as above, and $\lambda$, $P$, $Q$ are functions of $\theta$ and $\tau$.
The Einstein equations for (1.3) are a set of coupled nonlinear equations
for $P$ and $Q$, and, again, first-order equations for $\lambda$ that
can be integrated once $P$ and $Q$ are known.

The models (1.3) have been used for numerical studies [5], since it is
usually assumed that there is little chance of finding analytic
solutions for such a complicated system of equations.  One aim of these
studies has been to show that, in general, the solutions to these
models become asymptotically velocity term dominated (AVTD) as $\tau
\rightarrow +\infty$, that is, near the singularity of the model.  The
idea of AVTD models, originally due to Belinskii, Khalatnikov and Lifshitz
[6] (BKL), is that near a singularity
in the ADM action,
$$I = {{1}\over {16\pi}} \int [\pi^{ij} \dot g_{ij} - N{\cal H}_{\perp}
- N_i {\cal H}^i]dt d^3x, \eqno (1.4)$$
$${\cal H}_{\perp} = {{1}\over {\sqrt{g}}}[\pi^{ij} \pi_{ij} - {{1}\over
{2}}(\pi^k_{\, k})^2] - \sqrt{g} ^3 \! R], \eqno (1.5a)$$
$${\cal H}^i = -2\pi^{ij}_{\, \, |j}, \eqno (1.5b)$$
where $^3 \! R$ is the curvature on $t$ = const. surfaces and $|$ is a
covariant derivative on these surfaces, the ${\cal H}^i = 0$ terms
will imply that the curvature (``potential'') term in the ${\cal H}_{\perp}$
can, at best, be ignored completely, or at worst become algebraic in
$g_{ij}$, so that the ADM equations can be solved as if the metric were
homogeneous (with the $g_{ij}$ completely independent functions of time
at each point $\bf x$).  The conjecture of BKL is that this behavior is
generic, an idea which is still being debated.  Numerical studies of the
models (1.3) (see [2] and references therein) have shown that these models
do indeed become AVTD except at a set of
isolated points ($\theta = \theta_i$),
where there are ``spikes'' in the solution.  The origin of these spikes
is studied in Ref. [2].

While the $T^3$ Gowdy models have been used extensively in classical
and minisuperspace quantum gravity [7], the $S^1 \times S^2$ topology seems
to have been almost ignored.  The polarized form of these models can be
written, in the notation of [2], as
$$ds^2 = e^{-\lambda /2} e^{\tau /2} (-e^{-2\tau} d\tau^2 + d\theta^2)$$
$$+ \sin (e^{-\tau}) [e^P d\delta^2 + e^{-P} \sin^2 \theta d\phi^2],
\eqno (1.6)$$
where $g_2 = \sin^2 T \sin^2 \theta$ is a solution of (1.2).  These models
have Einstein equations that consist, as above, of a linear equation for
$P$ and first-order equations for $\lambda$ that can be directly integrated.
The linear equation can be solved by separation of variables, and several
particular solutions were given by Gowdy [1].

The unpolarized form of these models,
$$ds^2 = e^{-\lambda /2} e^{\tau /2} (-e^{-2\tau} d\tau^2 + d\theta^2)$$
$$ + \sin (e^{-\tau}) [e^P d\delta^2 + 2e^P Q d\delta d\phi +(e^P Q^2 +
e^{-P} \sin^2 \theta ) d\phi^2], \eqno (1.7)$$
(where, once again, $g_2 = \sin^2 e^{-\tau} \sin^2 \theta$ gives the form
of the $d\phi^2$ term), has Einstein equations similar to those of (1.3),
though quite a bit more complicated.  These models seem to have been
ignored, and there are, as far as we know, no numerical or analytic
solutions given in the literature.

One could argue that the fact that the Einstein equations for (1.7) are
even more complicated than those for (1.3) would mean that only numerical
solutions could be found.  However, we will show that at least one family of
particular analytic solutions is readily available as a reinterpretation
of the portion of the Kerr metric [7] between the inner, $r_{-} = M -
\sqrt{M^2 - a^2}$, and the outer, $r_{+} = M + \sqrt{M^2 -a^2}$ horizons
(where the ``radial'' coordinate is timelike).  We will study some of
the properties of this family of solutions and compare some of the
results with those of numerical solutions of (1.3). One important point
is that these solutions are {\it not\/} AVTD.  Of course, since such
particular solutions form a set of measure zero in the space of all solutions,
it is not clear what the generic behavior of these models near the
singularities is.  Another point is that the ``singularities'' of these
models are not true curvature singularities, since they represent points
where the solution crosses a lightlike surface into regions where the metric
is the usual stationary axisymmetric black hole metric.  We will discuss
possible generalizations briefly.

The paper is organized as follows:  in Sec. II we will give the Einstein
equations for all the metrics listed above, and show that the Kerr metric
between its horizons is a family of exact solutions for the unpolarized
$S^1 \times S^2$ model.  Sec. III discusses the properties of the solutions
(including possible AVTD behavior).  Sec. IV consists of conclusions and
some suggestions for extending the results of this article.
\vfill\eject

\noindent {\bf II. Exact Solutions}
\vskip 10 pt

\noindent {\it Equations of Motion\/}

The equations of motion for the polarized Gowdy models were given
in the original works of Gowdy [1, 4].  In our parametrization the equations
for the $T^3$ cosmologies become, as we have mentioned above, a linear
equation for $P$,
$$P_{,\tau \tau} - e^{-2\tau}P_{,\theta \theta} = 0, \eqno (2.1)$$
and a set of first-order equations for $\lambda$,
$$\lambda_{,\tau} = P_{,\tau}^{\, 2} + e^{-2\tau} P_{,\theta}^{\, 2},
\eqno (2.2a)$$
$$\lambda_{,\theta} = 2P_{,\theta} P_{,\tau}, \eqno (2.2b)$$
which, in principle can be integrated directly to give $\lambda$ if they
obey an integrability condition which is just (2.1).  Eq. (2.1) is
exactly soluble in terms of trigonometric functions of $\theta$ and
ordinary Bessel functions of $\tau$, and there are even tabulated integrals
that give $\lambda$ explicitly in certain cases [8].  Some of these
solutions were given by Gowdy in his original papers.

The equations for models with $S^1 \times S^2$ topology are only slightly
more complicated, again we have a linear equation for $P$,
$$P_{,\tau \tau} - {{e^{-2\tau}}\over {\sin \theta}} (\sin \theta
P_{, \theta})_{,\theta} - e^{-2\tau} -
[e^{-\tau} \cot (e^{-\tau}) - 1] P_{,\tau} =
0, \eqno (2.3)$$
this equation can be separated in terms of Legendre polynomials
and a function of $\tau$.  Some exact solutions to this
equation were given by Gowdy [1] (in another coordinate system).  The
equations for $\lambda$ become
$$\cot (e^{-\tau})\lambda_{,\theta} - 2e^{\tau} (P_{,\tau} P_{,\theta}) +
\cot \theta [-e^{\tau} \lambda_{,\tau} + 2e^{\tau} P_{,\tau} + e^{\tau}
+2\cot (e^{-\tau})] = 0, \eqno (2.4a)$$
$$\cot (e^{-\tau}) (\lambda_{,\tau} - 1) - e^{\tau} [(P_{,\tau})^2 +
e^{-2\tau}(P_{,\theta})^2 ]$$
$$+ e^{-\tau} [\cot^2 (e^{-\tau}) + 4] +
e^{-\tau} (-\cot \theta \lambda_{,\theta} + 2\cot \theta P_{,\theta} ) = 0.
\eqno (2.4b)$$

The unpolarized models have somewhat more complicated equations.  The
equations for $P$ become nonlinear because nonlinear terms in $Q$ are added,
and an equation for $Q$ appears which has nonlinear terms in both $Q$ and
$P$.  The $T^3$ model has been used extensively for numerical studies
of such concepts as AVTD models [2, 5].  Its equations for $P$
and $Q$ are
$$P_{,\tau \tau} - e^{-2\tau} P_{,\theta \theta} - e^{2P}(Q_{,\tau}^2 -
e^{-2\tau}Q_{,\theta}^2) = 0, \eqno (2.5a)$$
$$Q_{,\tau \tau} - e^{-2\tau}Q_{,\theta \theta} + 2(P_{,\tau}Q_{,\tau} -
e^{-2\tau} P_{,\theta} Q_{,\theta}) = 0, \eqno (2.5b)$$
and once these are solved one can intregrate the equations for $\lambda$,
$$\lambda_{,\tau} - [P_{,\tau}^2 + e^{-2\tau} P_{,\theta}^2 + e^{2P}(
Q_{,\tau}^2  + e^{-2\tau} Q_{,\theta}^2)] = 0, \eqno (2.6a)$$
$$\lambda_{,\theta} - 2(P_{,\theta} P_{,\tau} + e^{2P} Q_{,\theta} Q_{,\tau})
= 0. \eqno (2.6b)$$
One of the reasons usually given for solving these equations numerically
is that there is little chance of finding analytic solutions to such a
complicated system of nonlinear equations.

The unpolarized $S^1 \times S^2$ models have even more complicated
dynamical equations for $P$ and $Q$,
$$P_{,\tau \tau} - e^{-2\tau} {{(\sin \theta P_{,\theta})_{,\theta}}\over
{\sin \theta}} - e^{-2\tau} - {{e^{2P}}\over {\sin^2 \theta}}\{ (Q_{,\tau})^2
- e^{-2\tau} (Q_{,\theta})^2\} - [e^{\tau} \cot (e^{-\tau}) - 1]P_{,\tau} =
0, \eqno (2.7a)$$
$$Q_{,\tau \tau} - e^{-2\tau} Q_{,\theta \theta} +
e^{-2\tau}\cot \theta Q_{,\theta} + 2(P_{,\tau} Q_{,\tau} -
e^{-2\tau} P_{,\theta} Q_{,\theta})- [e^{\tau} \cot (e^{-\tau}) - 1]
Q_{,\tau} = 0. \eqno (2.7b)$$
There are two coupled first-order equations for $\lambda$ (which,
of course, can be reduced to separate equations for $\lambda_{,\tau}$ and
$\lambda_{,\theta}$, but give equations that are more complicated than
the originals).  They are
$$\cot (e^{-\tau}) \lambda_{,\theta} - 2 e^{\tau}(P_{,\tau} P_{,\theta} +
e^{2P} {{Q_{,\tau} Q_{,\theta}}\over {\sin^2 \theta}}) $$
$$+ \cot \theta [-e^{\tau} \lambda_{,\tau} + 2e^{\tau} P_{,\tau} + e^{\tau}
+ 2\cot (e^{-\tau})] = 0, \eqno (2.8a)$$
$$\cot (e^{-\tau}) (\lambda_{,\tau} - 1) - e^{\tau}[(P_{,\tau})^2 +
e^{-2\tau}(P_{,\theta})^2] - e^{\tau}{{e^{2P}}\over {\sin^2 \theta}}
[(Q_{,\tau})^2 + e^{-2\tau}(Q_{,\theta})^2]$$
$$+ e^{-\tau}[\cot^2 (e^{-\tau}) + 4] + e^{-\tau}(-\cot \theta
\lambda_{,\theta} + 2\cot \theta P_{,\theta}) = 0. \eqno (2.8b)$$
One could expect that it might be difficult to find exact solutions to this
system of equations, but one is at hand without any work.
\vskip 10 pt

\noindent {\it Black Holes inside their Horizons.\/}
\vskip 10 pt

Kantowski and Sachs [9] studied cosmological models with four-dimensional
groups of motion which had three-dimensional subgroups, and one special case
was a cosmological model of the form
$$ds^2 = -N^2(t) dt^2 + e^{2\sqrt{3} \beta (t)} dr^2 + e^{-2\sqrt{3} \beta
(t)} e^{-2\sqrt{3} \Omega (t)} (d\theta^2 + \sin^2 \theta d\varphi^2).
\eqno (2.9)$$
They gave a one parameter family of solutions of the form
$$N^2 = {{1}\over {{{\alpha}\over {t}} - 1}}, \qquad e^{2\sqrt{3} \beta}
= {{\alpha}\over {t}} - 1, \qquad e^{-2\sqrt{3} \beta (t)} e^{-2\sqrt{3}
\Omega (t)} = t^2. \eqno (2.10)$$
They note that this family is nothing more than the Schwarzschild model
inside the horizon.  That is, if we write the usual Schwarzschild metric
in Schwarzschild coordinates, we have
$$ds^2 = -(1 - {{2m}\over {r}}) dt^2 + {{1}\over {1 - {{2m}\over {r}}}}dr^2
+ r^2 (d \theta^2 + \sin^2 \theta d\varphi ^2), \eqno (2.11)$$
and for $r < 2m$ we can make the coordinate transformation $t \leftrightarrow
r$ and we find (2.9) with the functions given by (2.10).

One advantage of looking at the Kantowski-Sachs model as Schwarzschild
is that we can see the character of the singular points of the metric.
The singularity at $t = 0$ is the true singular point of Schwarzschild
($r = 0$), where the curvature is infinite, but the singularity at
$t = \alpha$ is just a lightlike surface where the curvature is regular
and we pass from the Kantowski-Sachs region to the Schwarzschild region.
Something similar happens with the vacuum Taub
model [10], where there are two
singular points, but both of them are lightlike surfaces where we pass
into the NUT [11, 12] region.

The Gowdy models have two commuting Killing vectors, and in the $T^3$ and
$S^1 \times S^2$ topologies are axisymmetric.  If we were able to find a
static or stationary axisymmetric metric with a horizon of the Schwarzschild
type for which the region inside of the horizon had the proper character,
then one could find a solution to a Gowdy model in the same way that one
finds the Kantowski-Sachs solution from
Schwarzschild.  One difference is that in the Schwarzschild
case the Birkhoff theorem says that the Kantowski-Sachs model
is the general solution to the Einstein equations for this model.  In the
Gowdy case we can probably expect no more than a particular solution.

A metric that generalizes Schwarzschild and has two commuting Killing
vectors is the Kerr metric.  In Boyer-Lindquist coordinates the Kerr
metric is
$$ds^2 = -{{r^2 -2Mr + a^2}\over {r^2 + a^2 \cos^2 \theta}}[dt - a\sin^2
\theta d\phi]^2 + {{\sin^2 \theta}\over {r^2 + a^2\cos^2 \theta}}
[(r^2 + a^2)d\phi -adt]^2$$
$$+ {{r^2 + a^2 \cos^2 \theta}\over {r^2 -2Mr + a^2}}dr^2 + (r^2 + a^2 \cos^2
\theta)d\theta^2. \eqno (2.11)$$
For this metric the outer horizon, $r_{+} = M + \sqrt{M^2 - a^2}$ is a
lightlike surface where the light cones tip over to the point where
$\partial /\partial r$ is a timelike vector, so that a coordinate change
of the $r \leftrightarrow t$ is possible.  Unfortunately, at the inner
horizon, $r_{-} = M^2 - \sqrt{M^2 - a^2}$, another lightlike surface, the
light cones tilt back to the point where $\partial /\partial r$ becomes
spacelike again.  In the region $r_{-} < r < r_{+}$ we can make the
transformation $r \leftrightarrow t$ and the Kerr metric becomes a
cosmological model,
$$ds^2 = {{2Mt - t^2 - a^2}\over {t^2 + a^2 \cos^2 \theta}}[dr -
a^2 \sin^2 \theta d\phi]^2 + {{\sin^2 \theta}\over {t^2 + a^2 \cos^2 \theta}}
[(t^2 + a^2)d\phi - a dr]^2$$
$$- {{t^2 + a^2 \cos^2 \theta}\over {2Mt - t^2 - a^2}}dt^2 + (t^2 + a^2
\cos^2 \theta)d\theta^2. \eqno (2.12)$$
This metric, with the simple transformation
$$t = \alpha[\sqrt{1 - \beta^2}\cos(e^{-\tau}) + 1], \eqno (2.13)$$
$\alpha = M$, $\beta = a/M$, makes (2.12) into (1.7) with
$$\lambda = \tau - 2\ln (\alpha^2 [\sqrt{1 - \beta^2}\cos(e^{-\tau}) + 1]^2
+\beta^2 \cos^2 \theta), \eqno (2.14a)$$
$$P = \ln [(1 - \beta^2)\sin^2 (e^{-\tau}) + \beta^2 \sin^2 \theta] -
\ln[\alpha \sqrt{1 - \beta^2}\sin (e^{-\tau})]$$
$$-\ln ([\sqrt{1 - \beta^2}\cos(e^{-\tau}) + 1]^2 + \beta^2 \cos^2 \theta),
\eqno (2.14b)$$
$$Q = - {{2\alpha \beta \sin^2 \theta [\sqrt{1 - \beta^2}\cos (e^{-\tau})
+ 1]}\over {(1 - \beta^2)\sin^2 (e^{-\tau}) + \beta^2 \sin^2 \theta}}
\eqno (2.14c)$$
Even though it is obvious that Eqs. (2.14) are a solution of the vacuum
Einstein equations (since the Kerr metric is a solution), we have, using
a symbolic manipulation program, inserted (2.14)  into (2.7) and (2.8)
and showed explicitly that these expressions are solutions.  In the next
section we will discuss the properties of this family of solutions.
\vfill\eject

\noindent {\bf III. Properties of the Solutions}
\vskip 10 pt

The general behavior of this family of solutions is given in Figs. 1 and 2,
which show $P + \ln [\alpha \sqrt{1 - \beta^2} \sin (e^{-\tau})]$
and $Q$ as functions of $\theta$ for various values of $\tau$. The functions
are relatively flat at $\tau = -\ln \pi$, which is the ``big bang'', then
grow a spike in the region of $\theta = \pi /2$ which disappears as
$\tau \rightarrow +\infty$, so the functions become flat again at the
``big crunch''.  As $\tau$ goes either to $-\ln \pi$ or $+\infty$, $P$
becomes infinite, but the combination $P + \ln [\alpha \sqrt{1 - \beta^2}
\sin (e^{-\tau})]$ (i.e. adding a $\theta$-independent term) is finite
for all $\theta$ and $\tau$ except at the extremes of $\theta$, $\theta =0$
and $\theta = \pi$, where, for $\tau = -\ln \pi$ and $\tau = +\infty$,
it becomes $-\infty$ ($e^P = 0$).

At first glance the ``spiky'' behavior of these solutions seems to have
something to do with similar spikes that occur in numerical solutions.
While we will show that there is a remote connection between the two
situations, the spikes in our solution show up at different parts of the
evolution from the numerical spikes.

The simplest manner of showing the difference between the numerical and
analytic solutions is to study the Hamiltonian formulation of the
Einstein equations for the $S^1 \times S^2$ topology, which is similar to
that of the $T^3$ topology given in Refs. [2, 13].  Equations (2.7) can
be modified by defining $e^{P^{\prime}} = e^P/\sin \theta$ aand collecting the
first and last terms of each equation.  They become
$$e^{-\tau} \csc (e^{-\tau})(e^{\tau} \sin (e^{-\tau})P^{\prime}_{,\tau}
)_{,\tau} - e^{-2\tau}{{(\sin \theta P^{\prime}_{,\theta})_{,\theta}}\over
{\sin \theta}}$$
$$-e^{P^{\prime}}\{Q^2_{,\tau} + e^{-2\tau}Q^2_{,\theta}\} = 0, \eqno (3.1a)$$
$$e^{-\tau} \csc (e^{-\tau})(e^{\tau} \sin (e^{-\tau}) Q_{,\tau})_{,\tau}
- e^{-2\tau}{{(\sin \theta Q_{,\theta})_{,\theta}}\over {\sin \theta}}$$
$$+ 2(P^{\prime}_{,\tau} Q_{,\tau} - e^{-2\tau} P^{\prime}_{,\theta}
Q_{,\theta}) = 0. \eqno (3.1b)$$
If we define a new time variable $w$ by means of $dw = e^{-\tau} \csc
(e^{-\tau}) d\tau$, or
$$w = -\ln \left [\tan \left ( {{e^{-\tau}}\over {2}} \right ) \right ],
\eqno (3.2)$$
then, using $\sin (e^{-\tau}) = \, {\rm sech}\, w$, and multiplying
through by $e^{2\tau} \sin^2 (e^{-\tau})$, Eqs. (3.1) reduce to
$$P^{\prime}_{,ww} - \, {\rm sech}^2 \, w {{(\sin \theta P^{\prime}_{,\theta}
)_{,\theta}}\over {\sin \theta}} - e^{2P^{\prime}} \{Q^2_{,w} + \,
{\rm sech}^2 \, w \, Q^2_{,\theta}\} = 0, \eqno (3.3a)$$
$$Q_{,ww} - \, {\rm sech}^2 \, w{{(\sin \theta Q_{,\theta})_{,\theta}}\over
{\sin \theta}} + 2(P^{\prime}_{,w} Q_{,w} - \, {\rm sech}^2 \, w \,
P^{\prime}_{,\theta} Q_{,\theta}) = 0. \eqno (3.3b)$$
These equations are similar to those of the $T^3$ topology [2], with
$P_{,\theta \theta}$ and $Q_{,\theta \theta}$ replaced by $(\sin \theta
P^{\prime}_{,\theta})_{,\theta} /\sin \theta$, $(\sin \theta
Q_{,\theta})_{,\theta} /\sin \theta$ and $e^{-2\tau}$ replaced by
${\rm sech}^2 \, w$.  If we look at (3.2), we see that near the big bang,
$w \rightarrow -\infty$, and near the big crunch, $w \rightarrow +\infty$,
${\rm sech}^2 \, w \rightarrow e^{-2|w|}$.

Given the similarity between Eqs. (3.3) and those of the $T^3$ case, it is
not difficult to construct the Hamiltonian form of the action that gives
them [14],
$$I = \int d\theta \, dw \bigg [\Pi_{P^{\prime}} P^{\prime}_{,w} +
\Pi_Q Q_{,w}$$
$$ - {{1}\over {2}} \left \{ {{\Pi^2_{P^{\prime}}}\over {\sin \theta}} +
{{e^{-2P^{\prime}} \Pi^2_Q}\over {\sin \theta}} + \sin \theta \,
{\rm sech}^2 \, w [(P^{\prime}_{,\theta})^2 + e^{2P^{\prime}}Q^2_{,\theta}]
\right \} \bigg ]. \eqno (3.4)$$
Since ${\rm sech}^2 \, w \rightarrow e^{-2|w|}$ for $w \rightarrow
\pm \infty$, one might expect that the solutions (2.14) would become AVTD
at both the big bang and the big crunch.  In this case the potential terms
would become negligible and the Hamiltonian density would approach
$${\cal H} = {{\Pi^2_{P^{\prime}}}\over {2\sin \theta}} +
{{e^{-2P^{\prime}} \Pi^2_Q}\over {2\sin \theta}}, \eqno (3.5)$$
which is, again, similar to the $T^3$
expression given in Refs. [2, 13, 14]. This reduced
Hamiltonian has an exact solution of the form
$$P^{\prime} = P_0 + \ln [\cosh vw + \cos \psi \sinh vw ], \eqno (3.6a)$$
$$Q = Q_0 + {{e^{P_0} \sin \psi \tanh vw}\over {1 + \cos \psi \sinh vw}},
\eqno (3.6b)$$
$$\Pi_{P^{\prime}} = \left ({{v\sinh vw + v\cos \psi \cosh vw}\over
{\cosh vw + \cos \psi \sinh vw}}\right ) \sin \theta, \eqno (3.6c)$$
$$\Pi_q = e^{P_0}v \sin \psi \sin \theta, \eqno (3.6d)$$
where $P_0$, $v$, $Q_0$ and $\psi$ are arbitrary functions of $\theta$.

If the solution (2.14) is to be AVTD at either the big bang or the big
crunch, we have to rewrite $P$ and $Q$ in terms of $w$ and see whether
$P^{\prime}$ and $Q$ take the form of $P^{\prime}$ and $Q$ in (3.6).
Since $P$ and $Q$ in (2.14) are functions of $\sin (e^{-\tau})$ and
$\cos (e^{-\tau})$, and we have $\sin (e^{-\tau}) = {\rm sech}\, w$,
$\cos (e^{-\tau}) = \tanh w$ and $P^{\prime} = P - \ln (\sin \theta)$, we
find that
$$P^{\prime} = \ln \left [{{(1 - \beta^2)\, {\rm sech}\, w + \beta^2 \sin^2
\theta \cosh w}\over {\alpha \sqrt{1 - \beta^2} \sin \theta ([
\sqrt{1 - \beta^2} \tanh w + 1]^2 + \beta^2 \cos^2 \theta)}} \right ],
\eqno (3.7a)$$
$$Q = {{-2\alpha \beta \sin^2 \theta [\sqrt{1 - \beta^2} \tanh w + 1]}\over
{(1 - \beta^2) \, {\rm sech}^2 \, w + \beta^2 \sin^2 \theta}}. \eqno (3.7b)$$
There are at least two ways to take the limit as $w \rightarrow \pm \infty$;
one is to assume that ${\rm sech}\, w \rightarrow 0$ and that
$$P^{\prime} \rightarrow \ln \left [ {{\beta^2 \sin \theta \cosh w}\over
{\alpha \sqrt{1 - \beta^2} (2 - \beta^2 \sin^2 \theta + 2\sqrt{1 -
\beta^2})}} \right ]$$
$$= \ln \left ({{\beta^2 \sin \theta \cosh w}\over {\alpha \sqrt{1 -
\beta^2} (2 - \beta^2 \sin^2 \theta + 2\sqrt{1 - \beta^2})}} \right )
+ \ln \cosh w, \eqno (3.8a)$$
$$Q \rightarrow -{{2\alpha}\over {\beta}}[1 + \sqrt{1 - \beta^2} \tanh w],
\eqno (3.8b)$$
which imply $v = 1$, $\psi = \pi /2$, $P_0 = \ln (\beta \sin \theta/
\alpha \sqrt{1 - \beta^2}[2 - \beta^2 \sin^2 \theta + 2\sqrt{1 -
\beta^2}])$, $Q_0 = -2\alpha / \beta$. However, the coefficient of $\tanh w$
in (3.8b) is not $e^{-P_0}$, which it should be from (3.6b).  It is
worse if we calculate the momenta.  For example, $\Pi_Q = e^{P^{\prime}}
Q_{,w} \sin \theta$, and using the expression (3.7b) we find that $\Pi_Q$
calculated in this manner is not the same as (3.6d).

While this is the simplest limit, it is possible to keep terms of order
$e^{-w}$ and still have a limit that has approximately the form of (3.6).
Since ${\rm sech} \, w \rightarrow e^{-|w|}$, $\cosh w \rightarrow e^{+|w|}$,
and $e^{\pm|w|} = \cosh |w| \pm \sinh |w|$, and $\tanh w \rightarrow \pm 1$,
the limit of $P^{\prime}$ as $w \rightarrow \pm \infty$ is
$$P^{\prime} \rightarrow -\ln \left [{{\alpha \sqrt{1 - \beta^2}(2 -
\beta^2 \sin^2 \theta \pm 2\sqrt{1 - \beta^2})}\over {1 - \beta^2 \cos^2
\theta}}\right ]$$
$$+ \ln \left [ \cosh w \mp {{[1 - \beta^2(1 + \sin^2 \theta)]}\over
{1 - \beta^2 \cos^2 \theta}} \sinh w \right ], \eqno (3.9)$$
which implies $\cos \psi = \mp [1 - \beta^2(1 + \sin^2 \theta)]/ [
1 - \beta^2 \cos^2 \theta]$, $v = 1$, \hfil\break
$P_0 = (1 - \beta^2 \cos^2 \theta)/
\{\alpha \sqrt{1 - \beta^2}(2 - \beta^2 \sin^2 \theta \pm 2\sqrt{1 -
\beta^2})\}$.

As $w \rightarrow \pm \infty$, $\tanh w \rightarrow \pm (1 - {{1}\over {2}}
e^{-2|w|})$ and $Q$ becomes
$${{-2\alpha \beta \sin^2 \theta [\pm \sqrt{1 - \beta^2}(1 - {{1}\over {2}}
e^{-2|w|}) + 1]}\over {(1 - \beta^2) e^{-2|w|} + \beta^2 \sin^2 \theta}},
\eqno (3.10)$$
and using, as above, $e^{\pm |w|} = \cosh |w| \pm \sinh |w|$, we find that
$$Q \rightarrow {{-2\alpha \beta \sin^2 \theta [(1 + {{1}\over {2}}
\sqrt{1 - \beta^2})\cosh w \pm (1 + {{3}\over {2}}\sqrt{1 - \beta^2} \sinh
w]}\over {(1 - \beta^2 \cos^2 \theta)\cosh w \pm [-1 + \beta^2 (1 + \sin^2
\theta)]
\sinh w}}$$
$$= {{-2 \alpha \beta \sin^2 \theta (1 + {{3}\over {2}}\sqrt{1 - \beta^2})
\left [{{1 + 1/2\sqrt{1 - \beta^2}}\over {1 + 3/2\sqrt{1 - \beta^2}}} \pm
\tanh w\right ]}\over {(1 - \beta^2 \cos^2 \theta)[\left [ 1 \mp
{{\beta^2 (1 + \sin^2 \theta) - 1}\over {1 - \beta^2 \cos^2 \theta}} \tanh w
\right ]}}, \eqno (3.11)$$
which is of the form of (3.6b) with $Q_0 = -2\alpha \beta \sin^2 \theta
(1 + {{1}\over {2}}\sqrt{1 - \beta^2})/(1 - \beta^2 \cos^2 \theta)$ and
$\psi$ as above.  However, if we calculate $\Pi_Q$ as before, it is not the
same as (3.6d).

The fact that these solutions do not become AVTD is, in fact, not surprising
if we study the velocity $v$ [2, 14],
$$v = {{1}\over {\sin \theta}}\sqrt{\Pi^2_{P^{\prime}} + e^{-2P^{\prime}}
\Pi^2_Q}$$
that appears in (3.6).  This quantity can be calculated for all values of
$w$, and is
$$v(\theta, w) = {{{\rm sech}\, w}\over {[(1 - \beta^2)\,{\rm sech}^2 \, w +
\beta^2 \sin^2 \theta]([\sqrt{1 - \beta^2}\tanh w + 1]^2 + \beta^2 \cos^2
\theta)}}$$
$$\times \bigg ( \{\sinh w[(1 - \beta^2)\, {\rm sech}^2\, w - \beta^2
\sin^2 \theta]([\sqrt{1 - \beta^2}\tanh w + 1]^2 + \beta^2 \cos^2 \theta)$$
$$+ 2\, {\rm sech}\, w[(1 - \beta^2)\tanh w + \sqrt{1 - \beta^2}][(1 -
\beta^2)\, {\rm sech}^2\, w + \beta^2 \sin^2 \theta]\}^2$$
$$+ 4\beta^2
\sin^2 \theta ([\sqrt{1 - \beta^2}\tanh w + 1]^2 - \beta^2 \cos^2 \theta)^2
\bigg )^{1/2}. \eqno (3.12)$$
As  $w \rightarrow \pm \infty$, $v \rightarrow 1$ and $P^{\prime} \rightarrow
|w|$.

If we consider Eq. (3.3a), the solution will be AVTD if the terms involving
$\theta$-derivatives decay much more rapidly than those involving
$w$-derivatives.  The term involving $Q^2_{,\theta}$,
$${\rm sech}^2 \, w e^{2P^{\prime}}Q^2_{,\theta} \rightarrow e^{-2|w|}
e^{2|w|}Q^2_{,\theta}$$
does not necessarily decay rapidly as $w \rightarrow \pm \infty$. However,
calculating $Q_{,\theta}$, we see that it goes to zero as $w \rightarrow
\pm \infty$, so this term does not contribute in the limit.  However,
calculating $e^{2P^{\prime}}Q^2_{,w}$ and $P^{\prime}_{,ww}$, we see that they
decay as ${\rm sech}^2\, w$ (which is consistent with the AVTD form from
[3.6], where they should decay as ${\rm sech}^2\, vw$).  However,
$(\sin \theta P^{\prime}_{,\theta})_{,\theta}/\sin \theta$ becomes constant
in $w$ as $w \rightarrow \pm \infty$, so ${\rm sech}^2\, w\, (\sin \theta
P^{\prime}_{,\theta})_{,\theta}/\sin \theta$ is of the same order as
$P^{\prime}_{,ww}$ and $e^{2P^{\prime}}Q^2_{,w}$, and it is not surprising
that the solution (2.14) does not become AVTD as $w \rightarrow \pm \infty$.
Of course, if $v$ were less than one, $P^{\prime}_{,ww}$ and $e^{2P^{\prime}}
Q^2_{,w}$ would decay more slowly than ${\rm sech}^2\, w \,(\sin \theta
P^{\prime}_{,\theta})_{,\theta}/\sin \theta$.

As we mentioned above, the ``spiky'' behavior occurs far from the big bang
and the big crunch.  From Figs. 1 and 2
we see that the maximum deviation from a
constant function of $\theta$ occurs at $\tau = -\ln (\pi/2)$, which is at
$w = 0$.  It is not difficult to show that the solution near $w = 0$
is not AVTD, but a formal investigation of the behavior of $v$ near $\theta
= \pi/2$ and $w = 0$ sheds some light on the origin of the spike at $\theta
= \pi/2$.

If we consider $v(\theta, 0)$, we find that
$$v(\theta, 0) = {{2\sqrt{1 - \beta^2 \cos^2 \theta}}\over {1 + \beta^2
\cos^2 \theta}} \eqno (3.13)$$
which varies from $2\sqrt{1 - \beta^2}/(1 + \beta^2)$ at $\theta = 0, \pi$ to
two at $\theta = \pi/2$.  The value at $\theta = 0, \pi$ is always less
than two, and for $\beta$ near one the difference becomes large.  Notice
that the system becomes {\it formally\/} AVTD at $w = 0$, $\theta = 0$, in
that both $(P^{\prime}_{,\theta})^2$ and $(Q_{,\theta})^2$ are zero at this
point, while $v = 2$.  If one could assume that this fact was relevant to
the time development of the system, it would show that there is an indication
that the spiky behavior of Refs. [2, 5], which is tied to $v > 1$ at specific
points of the manifold, is present here also.  However, in our case,
``velocity dominated'' at one point in $\theta$ is meaningless, since the
$\theta$-derivative terms in (3.3) can be shown to be considerable at
$\theta = \pi/2$.  Nevertheless, the existence of these velocity peaks may
give us some clue about the true meaning of the spikes.

One last property of our solutions is their behavior as functions of $\beta$.
As $\beta \rightarrow 0$, the Kerr solution (2.12) becomes Schwarzschild, and
our Gowdy model becomes Kantowski-Sachs (compactified in the ``radial''
coordinate), and one of the singularities ($w \rightarrow -\infty$) becomes
a curvature singularity.  For the equivalent of extreme Kerr, $\beta = 1$,
both $\lambda$ and $Q$ seem reasonable, although time independent, but $P$
is singular.  However, the inner and outer horizons become the same, and there
is no Gowdy solution.  Notice that $t$ from (2.13) becomes $\alpha$ for all
values of $\tau$.
\vfill\eject

\noindent {\bf IV. Conclusions and Suggestions for Further Research}
\vskip 10 pt

We have shown that the Kerr metric between the inner and outer horizons
can be interpreted as an exact solution to the Gowdy models with
$S^1 \times S^2$ topology.  The Gowdy solutions generated by this method
are relatively flat functions of $\theta$ at the horizons (the ``big
bang'' at the inner horizon and the ``big crunch'' at the outer horizon)
which grow a spike near $\theta = \pi /2$ in both the Gowdy functions
$P(\theta , \tau)$ and $Q(\theta , \tau)$.

The singularities in the solutions are just lightlike surfaces where the
solution passes from the coordinate patch suitable for the Gowdy
interpretation to patches where we have true Kerr (black hole) metrics.
One important point is that we have compactified in the $\delta$-direction
which becomes the $t$-direction in the Kerr patches, giving solutions
with closed timelike lines in the outer regions, a property shared with the
Taub-NUT solution.  Notice that if one compactifies the Kantowski-Sachs
model (2.9) in the $r$-direction, there are also closed timelike lines
in the Schwarzschild portion.

The solutions we have found are not AVTD near the singularities, which can be
shown directly from the solutions and the Hamiltonian form of the action
which generates their equations of motion.  The fact that the velocity
that would appear in the AVTD solution is one seems to give a convincing
argument why this is so.

There are a number of directions in which this analysis can be extended.
An obvious idea is to study numerical solutions of Eqs. (3.3), which might
shed light on the structure of solutions for these models more general than
the particular solution (2.14), and could give reasons for the form that
(2.14) takes.

Another possibility [16] would be to create new solutions to the models
where the big bang and big crunch ``singularities'' become true
curvature singularities.  The technique for doing this is described
in Ref [17].

One final way to extend our idea would be to consider other models which
have cosmological sectors and ``black hole'' sectors separated by horizons.
There are a number of metrics where one or the other of these sectors is
well known, but where the other has never been interpreted as either a
``black hole'' solution or a cosmological model.  The cosmological sector of
some of these metrics can be compactified as we did with the metric (1.7) (or
they may naturally admit a compact topology),
leading to ``black hole'' sectors
with closed timelike lines.  The global geometry of some of these
manifolds (including the Kantowski-Sachs model compactified in the $r$
coordinate of Eq. [2.9]) has not been investigated in detail.
\vskip 20 pt

\noindent {\bf Acknowledgements}
\vskip 10 pt

We would like to thank H. Quevedo and R. Sussman for help with using
symbolic manipulation programs to calculate the Einstein equations
for the Gowdy models and to show that our solutions are indeed solutions.
We also thank V. Moncrief and T. Jacobson for very helpful discussions.
This work was supported in part by CONACyT grant 28454-E and DGAPA-UNAM
grant IN106097.
\vfill\eject

\noindent References
\vskip 10 pt

\noindent \item {1.} R. Gowdy, Phys. Rev. Lett. {\bf 27}, 826 (1971).

\noindent \item {2.} B. Berger and D. Garfinkle, Phys Rev. D {\bf 57}, 4767
(1998).

\noindent \item {3.} J. Synge, {\it Relativity: The General Theory\/}
(North Holland, Amsterdam, 1960).

\noindent \item {4.} R. Gowdy, Ann. Phys. (N. Y.) {\bf 83}, 203 (1974).

\noindent \item {5.} B. Berger and V. Moncrief, Phys. Rev. D {\bf 48},
4676 (1993).

\noindent \item {6.} V. Belinskii, E. Lifshitz and I. Khalatnikov, Sov.
Phys. Usp. {\bf 13}, 745 (1971); Adv. Phys. {\bf 19}, 525 (1970); see
also D. Eardley, E. Liang and R. Sachs, J. Math. Phys. {\bf 13}, 99
(1972).

\noindent \item {7.} Ch. Charac and S. Malin, Phys. Rep. {\bf 76}, 79 (1981);
C. Misner, Phys. Rev. D {\bf 8}, 3271 (1973); B. Berger, Ann. Phys. (N. Y.)
{\bf 83}, 458 (1975); Phys. Rev. D {\bf 11}, 2770 (1975); see also Ref. [9]
and R. Matzner, M. Rosenbaum and M. Ryan, J. Math. Phys. {\bf 23}, 1984
(1982).

\noindent \item {8.} R. Kerr, Phys. Rev. Lett. {\bf 11}, 237 (1963);
see also S. Hawking and G. Ellis, {\it The Large Scale Structure of
Space-Time\/}(Cambridge, Cambridge, 1973).

\noindent \item {9.} M. Ryan, in {\it SILARG VI\/}, edited by M. Novello
(World Scientific, Singapore, 1988).

\noindent \item {10.} R. Kantowski and R. Sachs, J. Math. Phys. {\bf 7},
443 (1966).

\noindent \item {11.} A. Taub, Ann. Math. (N. Y.) {\bf 53}, 472 (1951).

\noindent \item {12.} E. Newman, L. Tamburino and T. Unti, J. Math. Phys.
{\bf 4}, 915 (1963).

\noindent \item {13.} C. Misner, J. Math. Phys. {\bf 4}, 921 (1963).

\noindent \item {14.} V. Moncrief, Ann. Phys. (N. Y.) {\bf 132}, 87 (1981).

\noindent \item {15.} B. Grubi\v si\'c and V. Moncrief, Phys. Rev. D
{\bf 47}, 2371 (1993).

\noindent \item {16.} V. Moncrief, Private communication.

\noindent \item {17.} V. Moncrief, Class. Q. Grav. {\bf 4}, 1555 (1987).
\vfill\eject

\noindent FIGURE CAPTIONS
\vskip 10 pt

\noindent Figure 1a-g.  These figures show the evolution of $P - h$,
$h = \ln[\alpha\sqrt{1 - \beta^2} \sin (e^{-\tau})]$, as a function of
$\theta$ for the solution given in Eqs. (2.14) for $\alpha = 1$ and
$\beta = 1/2$. Fig. 1a corresponds to $\tau = -1.1447$, Fig. 1b to $\tau =
-1.144$, Fig. 1c to $\tau = -1.1$, Fig. 1d to $\tau = -0.75$,
Fig. 1e to $\tau = -\ln(\pi/2)$, Fig. 1f to $\tau = +5$, and Fig.
1g to $\tau = +10$.
\vskip 10 pt

\noindent Figure 2a-d.  These figures show the evolution of $Q$ as a function
of $\theta$ for the solution given in Eqs. (2.14) for $\alpha = 1$ and
$\beta = 1/2$.  Fig. 2a corresponds to $\tau = -1.1439$, Fig. 2b to $\tau =
-1.1$, Fig. 2c to $\tau = -\ln(\pi/2)$, and Fig. 2d to $\tau = +5$.
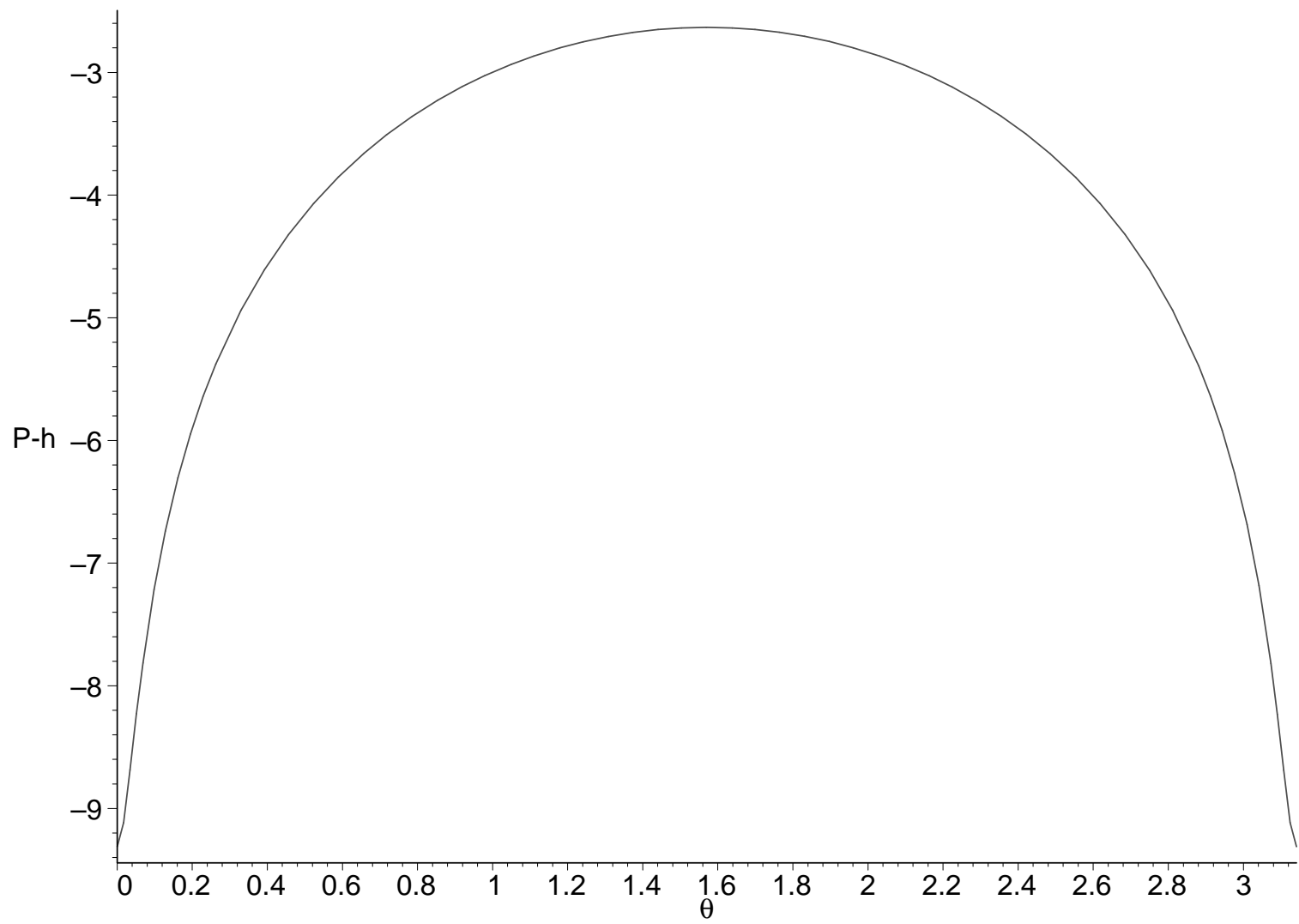
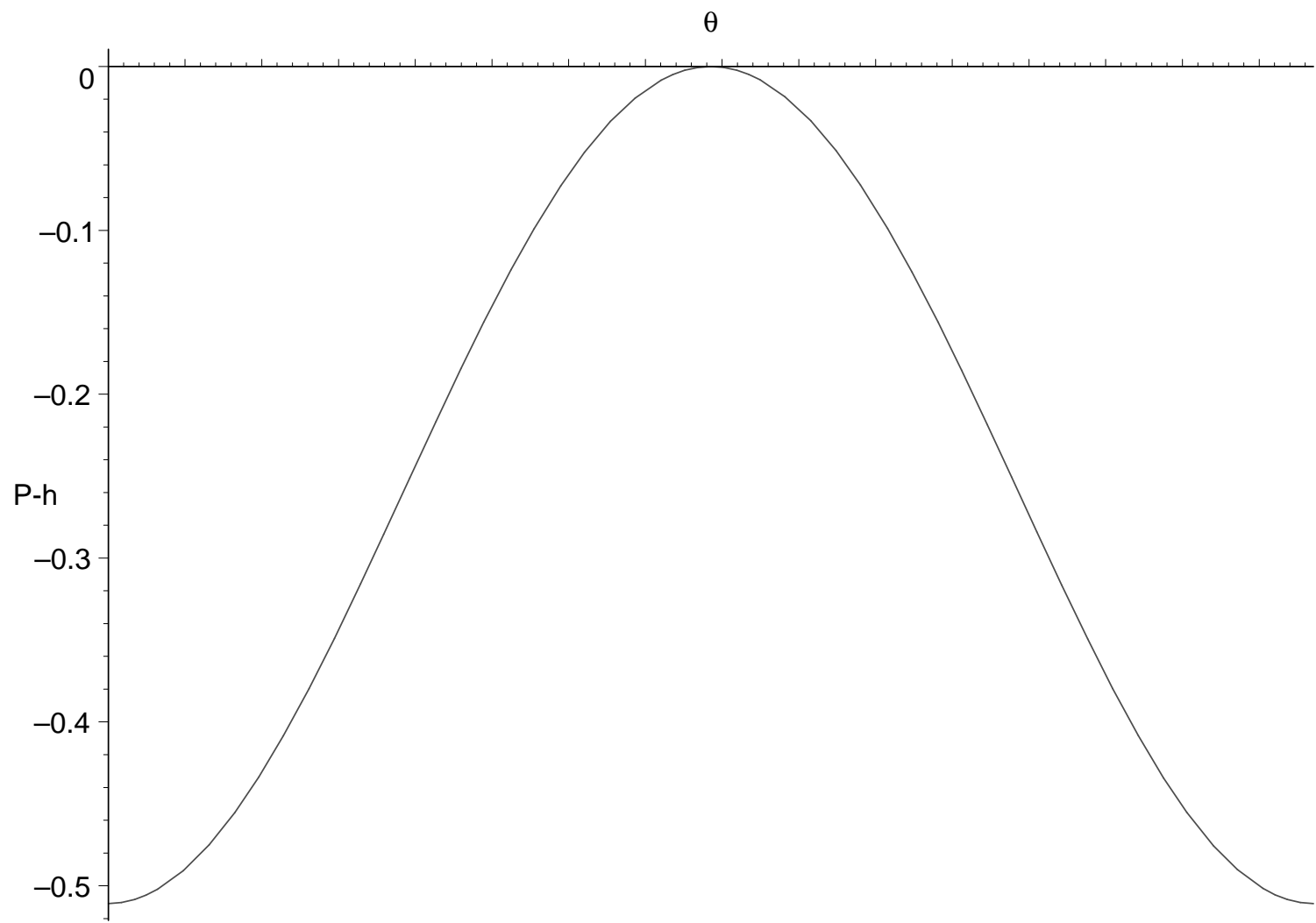
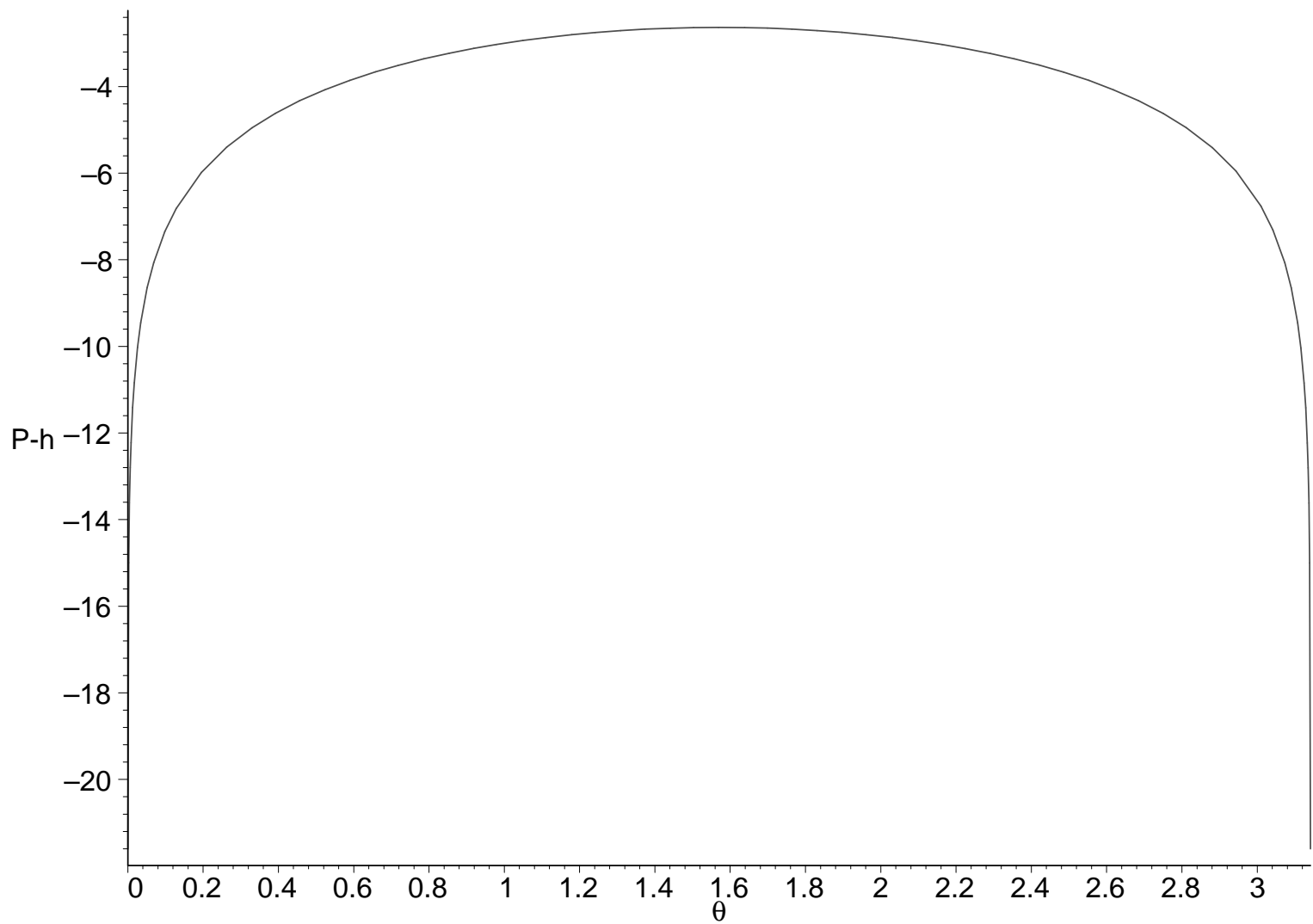
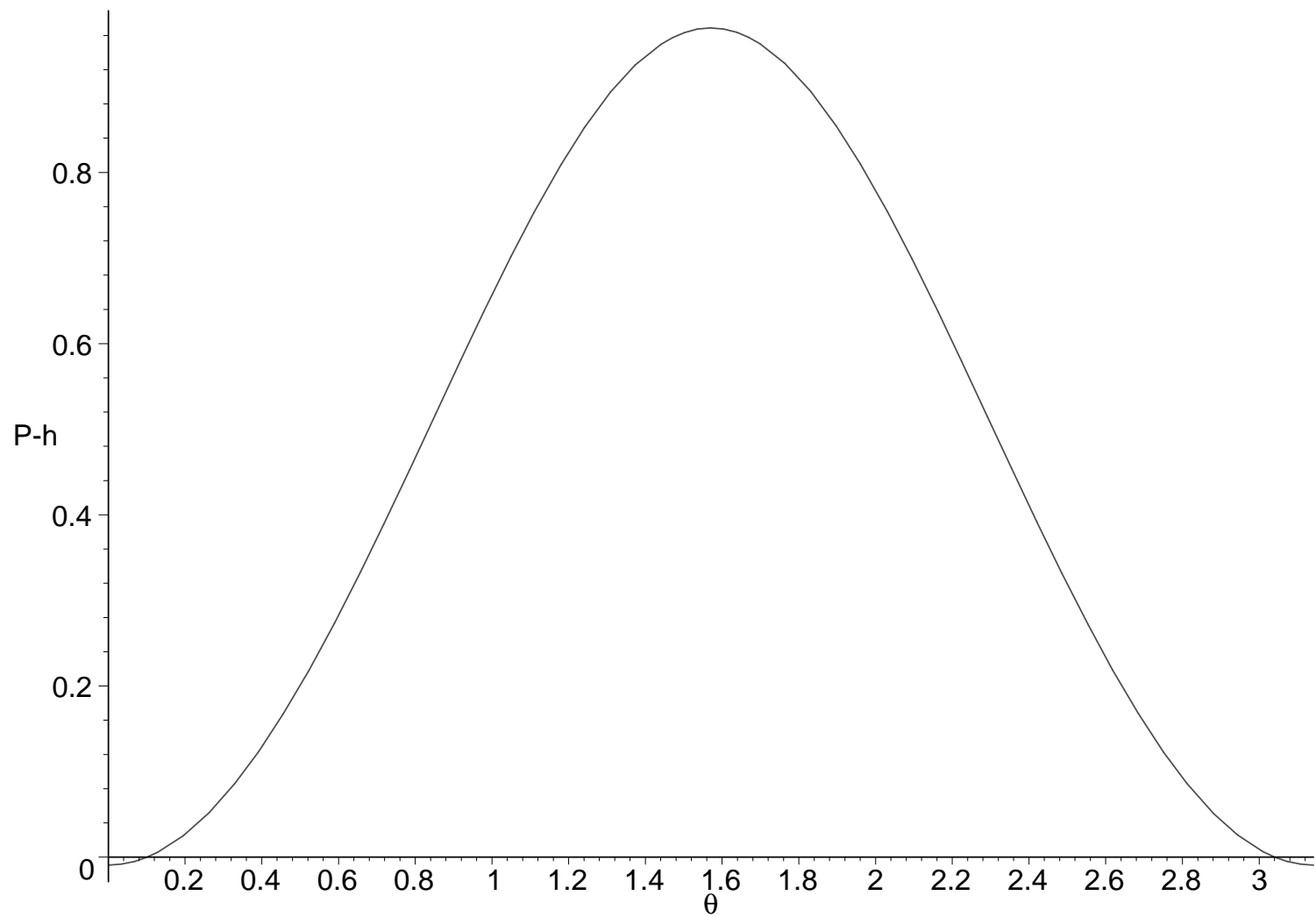
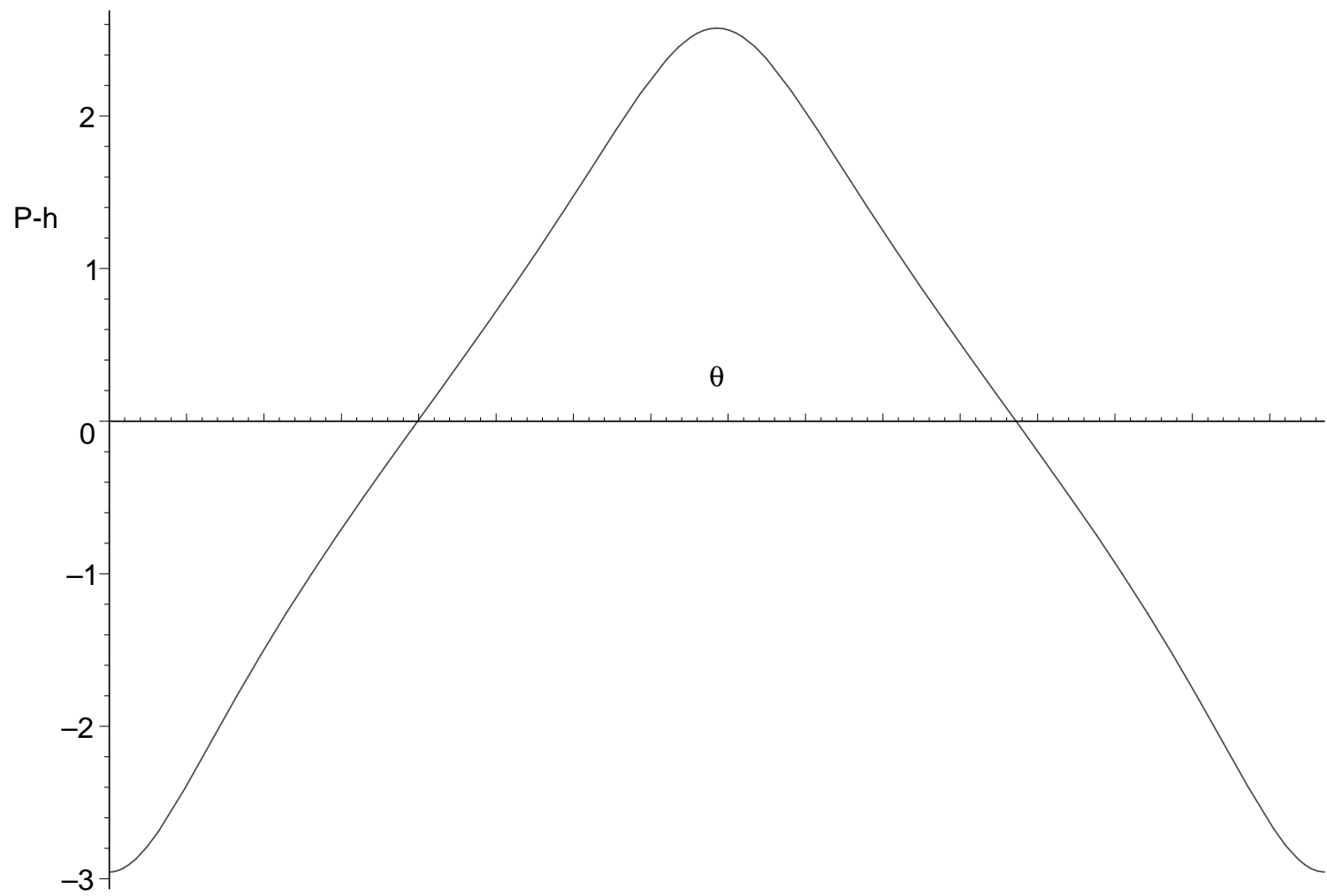
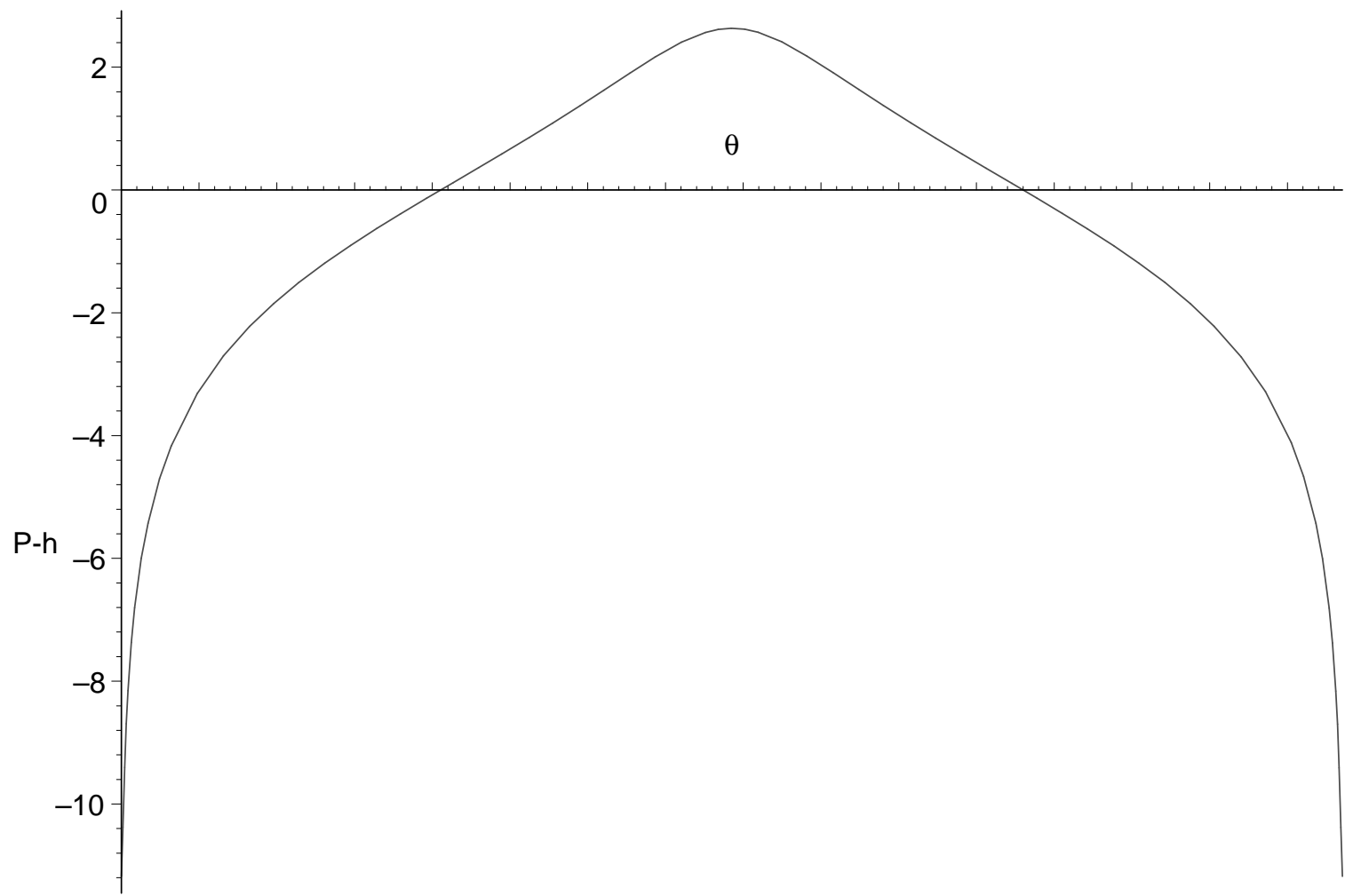
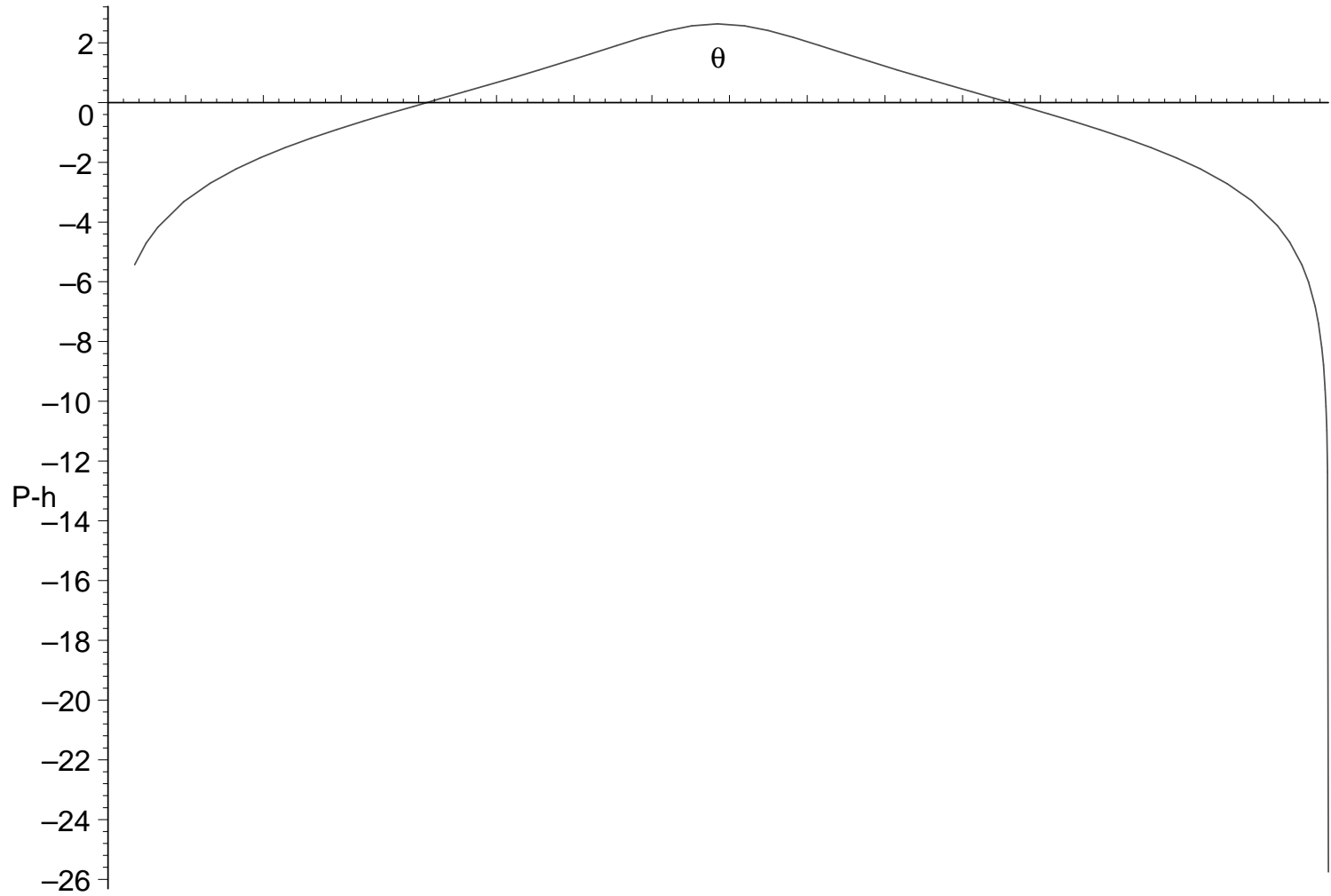
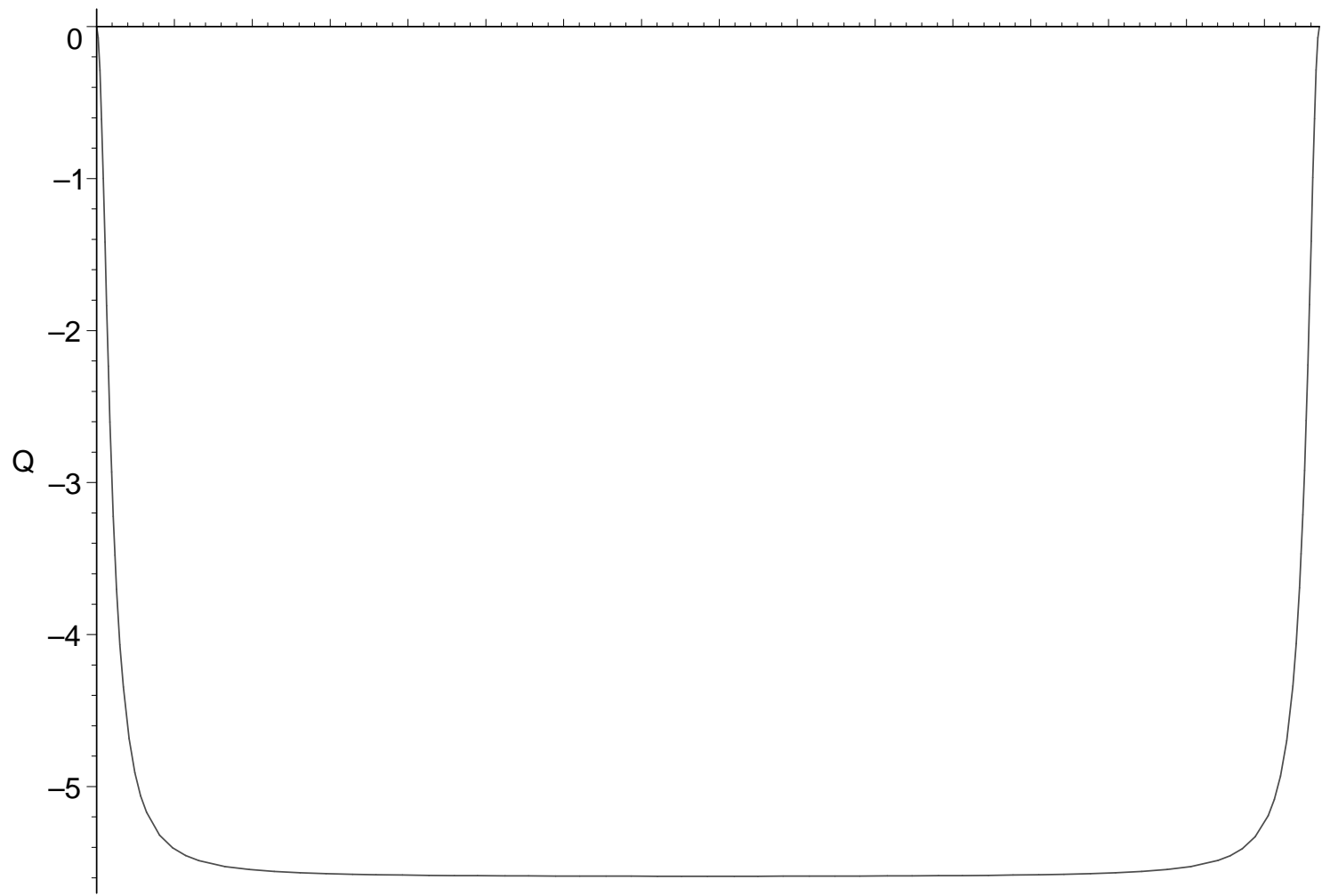
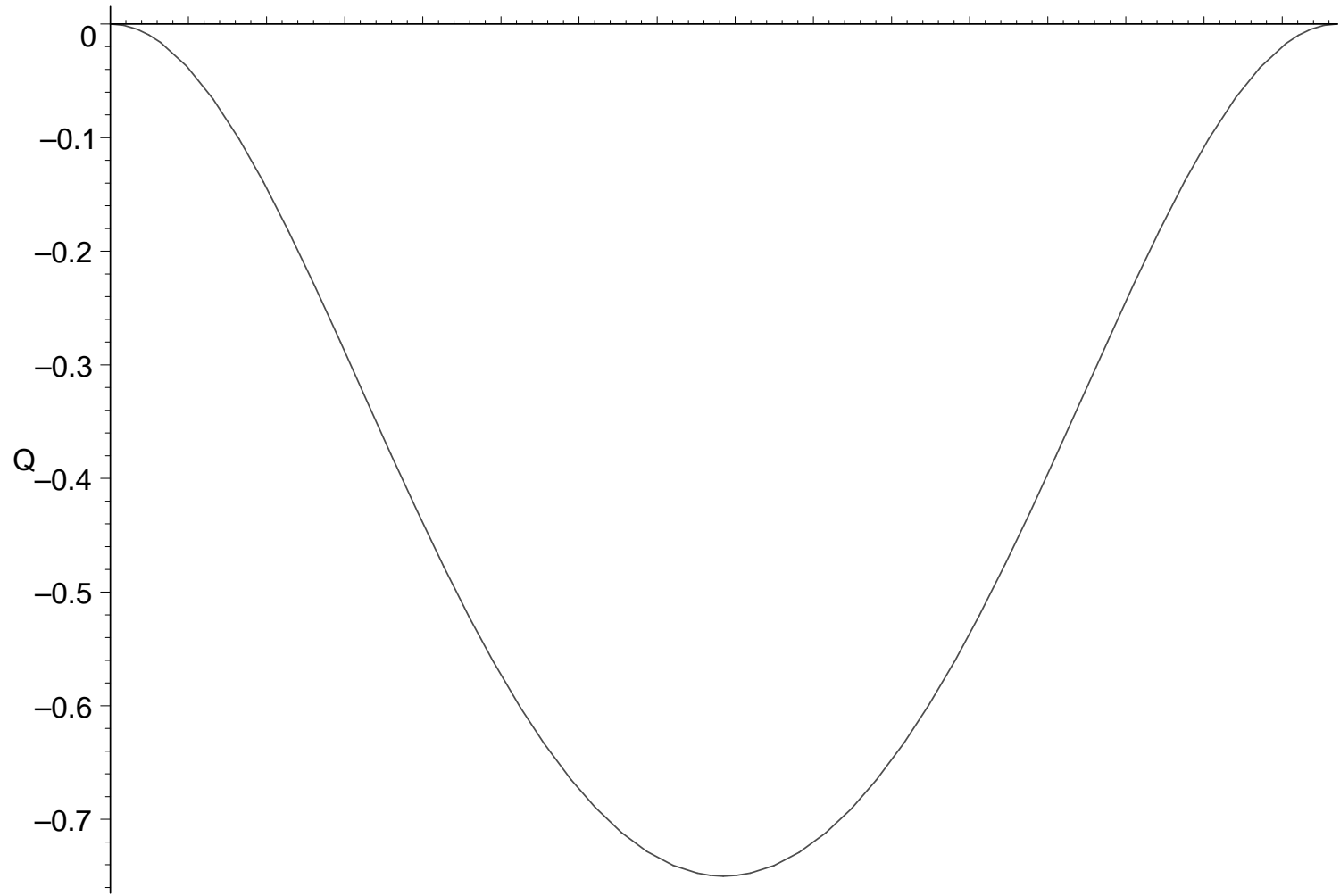
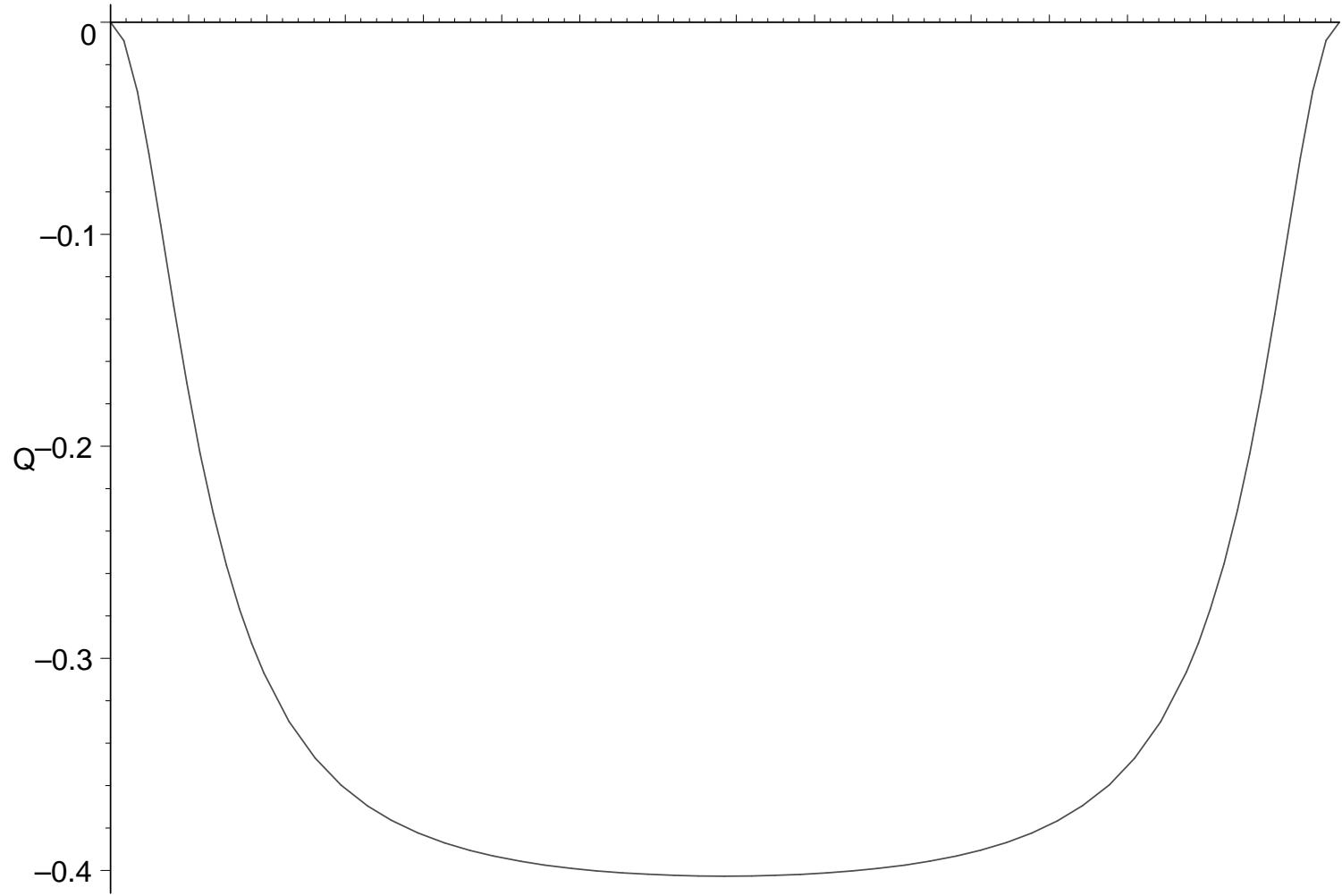
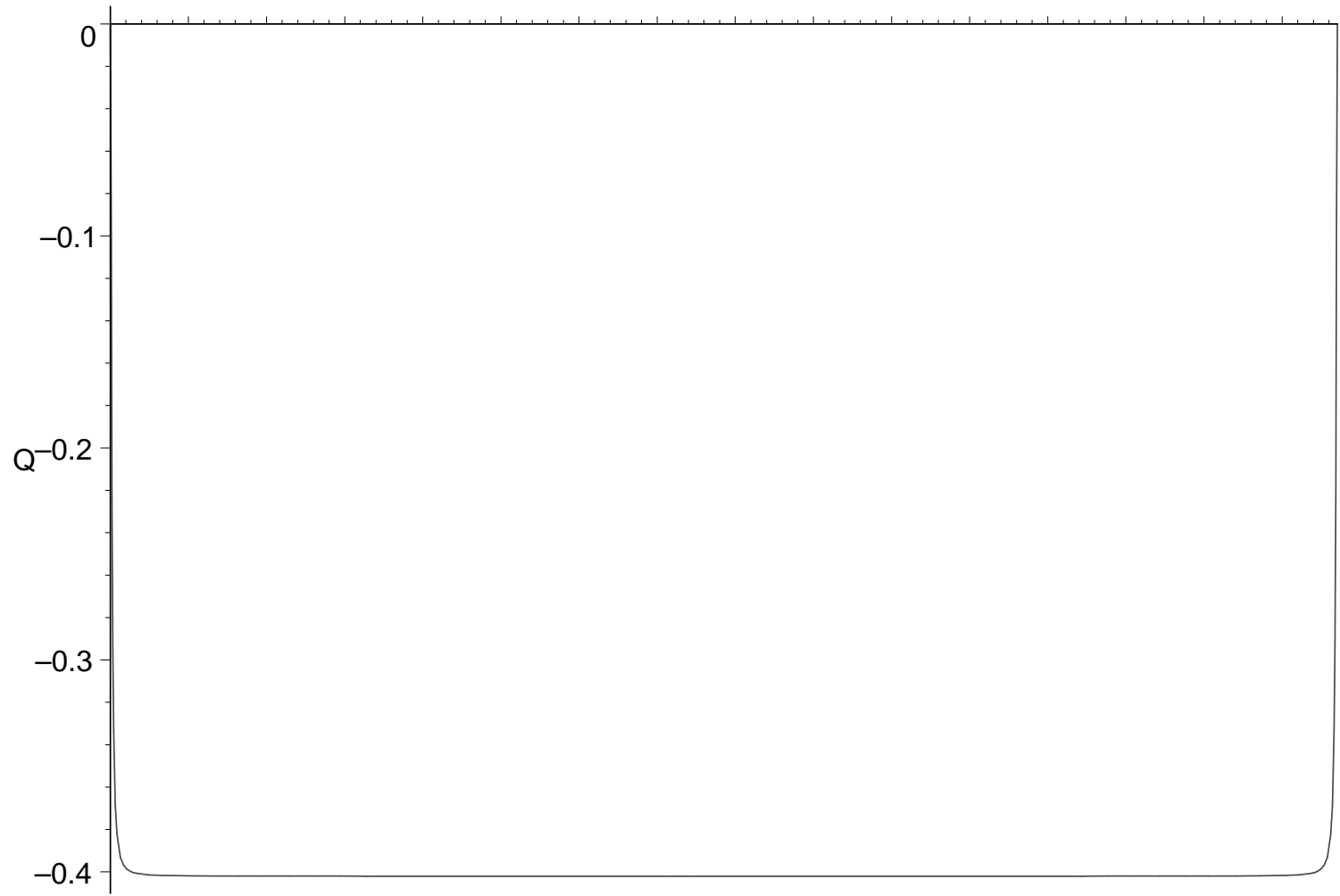
\end